\documentclass{svjour3}
\smartqed

\usepackage{babel}
\usepackage{graphicx}
\usepackage{color}
\usepackage{bm}
\usepackage{longtable}
\usepackage{amsmath,amsfonts,amssymb}
\usepackage{cite}
\usepackage{hyperref}
\usepackage{makecell,multirow}
\usepackage{tabularx}
\newcommand{\makeheadbox}{}
\usepackage{multicol}
\usepackage{array}
\usepackage{appendix}
\usepackage{subfig}

\begin{document}

\title{Near-perfect quantum teleportation between continuous and discrete encodings}

\author{Ravi Kamal Pandey$^{1}$, Shraddha Singh$^{2}$, Dhiraj Yadav$^{3}$, Devendra Kumar Mishra$^{1,*}$}

\institute{
$^{1}$Department of Physics, Institute of Science, Banaras Hindu University, Varanasi-221005, India\\
$^{2}$Department of Physics, Nehru Gram Bharti (Deemed to be University), Prayagraj-221505, India\\
$^{3}$Department of Physics, School of Sciences, IILM University, Greater Noida-201306, India\\
$^{*}$\email{kndmishra@gmail.com}}
\authorrunning{Ravi Kamal Pandey \textit{et. al}}
\maketitle

\abstract{
Quantum teleportation between polarized single-photon and phase-opposite coherent states is studied
using a hybrid entangled resource and entangled coherent states. The polarized single-photon qubit
represents a discrete-variable (DV) quantum system, whereas the phase-opposite coherent-state qubit
constitutes a continuous-variable (CV) system. While teleportation from CV to DV can be achieved
with near-unit success probability, the reverse process is usually limited to a maximum success
probability of $1/2$. We demonstrate that, by employing cross-Kerr nonlinearity together with passive
linear optical components such as polarizing beam splitters, beam splitters, and phase shifters,
almost perfect teleportation from DV to CV encodings can also be achieved.
}

\keywords{Quantum Teleportation \and Coherent States \and Hybrid Entanglement}

\section{Introduction}
Quantum teleportation \cite{bennett1993teleporting} is a method of transferring quantum state of a particle from one location to the other by using long range EPR \cite{einstein1935can} correlations and classical communication channel. Bennett et al. proposed a theoretical scheme of quantum teleportation for a two-level quantum system \cite{bennett1993teleporting} or more generally, for a DV quantum systems. The DV proposal for quantum teleportation was soon translated for a CV quantum systems \cite{vaidman1994teleportation}. Later, both DV and CV quantum teleportation as well as other quantum information protocols have been manifested in many experiments and in various possible systems\cite{bouwmeester1997experimental,boschi1998experimental,kim2001quantum,pirandola2015advances}. 

It has been one of the major concerns of the researchers to devise a robust physical system in which necessary manipulations in quantum states can be performed with convenience. In this regard, quantum optical states have proven to be of prime importance \cite{pirandola2015advances}. A qubit of information can be encoded in two orthogonal polarizations of a single photon, or, by using orthonormal states of vacuum and single photon states of an optical field. However, such DV optical systems are more vulnerable to losses in comparison to CV optical states like squeezed states and coherent states \cite{2010entangled}. The effect of decoherence on the quantum channel may be reversed using quantum repeaters and relays \cite{sangouard2011quantum} but there are some other restrictions with DV optical states as well. In the case of quantum teleportation of unknown DV state using linear optics, only two out of four Bell states can be discriminated with certainty, therefore, reducing success to $1/2$ \cite{lutkenhaus1999bell}. 

One of the alternative of DV optical systems is to encode information in the CV superposition of phase opposite coherent state of optical field \cite{glauber1963coherent}. In contrast to bi-photonic states, these are more robust against environmental decoherence. The other advantage is that almost perfect Bell state discrimination can be done using phase shifters, beam splitters and photon parity detectors which makes it an important resource for quantum teleportation as well as for other quantum communication tasks \cite{jeong2001quantum,wang2001quantum,jeong2002efficient,an2003teleportation,ralph2003quantum,
cheong2004near,prakash2009swapping,prakash2009entanglement,prakash2010improving
,mishra2010teleportation,sanders2012review,Prakash2019,pandey2019,pandey2021,pandey2022}. Van Enk et al. \cite{van2001entangled} studied quantum teleportation of such a superposition using entangled coherent state \cite{sanders1992entangled,sanders2012review} as resource,   reporting a success probability of $1/2$. The failure arises from the non-unitary nature of the operator $\hat{Z}_{c}=|\alpha\rangle\langle\alpha|-|-\alpha\rangle\langle-\alpha|$ required to be performed by the receiver for two cases of Bell state measurement. Many other quantum information processing tasks have been studied using these with similar value of success. Prakash et al. \cite{prakash2007improving} succeeded in improving the success probability to nearly unity for the quantum teleportation of entangled coherent state. The scheme only differs from the previous ones in differentiating the results of photon counting that the sender makes for Bell state measurement. Moreover, this scheme posed another practical difficulty by demanding the interconversion between even and odd coherent state as a part of unitary operation to be performed by the receiver. No one has shown how such a conversion would be done experimentally. 

The inference is that neither DV nor CV optical systems are capable of offering perfect quantum teleportation, each having its own merits and demerits. In order to accommodate the advantages of both, hybridization of DV and CV optical systems is important. In this regard, Lee et al. \cite{lee2013near} proposed a hybrid quantum information processing scheme using linear optics and hybrid entanglement of polarization and coherent state qubits. Park et al. \cite{park2012} discussed quantum teleportation between DV and CV quantum system encodings by using hybrid entangled resource in the presence of decoherence. Authors concluded that the success probability of teleportation depends on the direction of teleportation. It has been shown that the success probability from CV to DV becomes almost unity for significant coherent amplitude. However, in the case of teleportation from DV to CV system, the success reduces due to two reasons: the incapability of distinguishing the Bell states and, the non-unitary nature of $\hat{Z}$ needed to be performed by receiver on coherent state qubit. Therefore, the maximum success probability that can be obtained using their scheme is $1/2$. Later on, several other schemes for the generation of various hybrid entangled resource as well as for hybrid quantum information processing tasks have been proposed \cite{sheng2013,jeong2014,morin2014remote,jeong2016,sych2018,pod2019,
omkar2021,djor2022,bich2022}. 

We in our present work discuss a novel scheme for the quantum teleportation between a DV polarization qubit and a CV qubit encoded in phase opposite coherent states. A hybrid entangled resource is used for teleportation from CV to DV quantum system while a two mode entangled coherent state with appreciable large coherent amplitude is used for teleportation from DV to CV quantum system. We have shown that by using cross-Kerr nonlinear interaction \cite{he2011} and linear optical devices along with photon counting detectors, almost perfect teleportation in both directions can be obtained.
\section{Quantum teleportation from CV to DV encoding}
Let us assume that the sender, Alice, has a single qubit information,
\begin{equation}
\label{eqn:1}
|I\rangle_{0}=a|\alpha\rangle_{0}+b|-\alpha\rangle_{0}
\end{equation}
and she wish to teleport it to a distant partner Bob encoded in DV basis $\{|H\rangle,|V\rangle\}$. To achieve this task, Alice and Bob share modes 1 and 2, respectively, of the hybrid-entangled state,
\begin{equation}
\label{eqn:2}
|E\rangle_{1,2}=1/\sqrt{2}(|H,\alpha\rangle_{1,2}+|V,-\alpha\rangle_{1,2}).
\end{equation}
Alice mixes modes $0$ and $1$ in her possession using  a symmetric $50:50$ beam splitter to form output modes $3$ and $4$. The beam splitter which we abbreviate by $\mathcal{B}$, transforms input coherent states, $|{\alpha},{\beta}\rangle_{x,y}\rightarrow |\frac{{\alpha}+{\beta}}{\sqrt{2}},\frac{{\alpha}-{\beta}}{\sqrt{2}}\rangle_{u,v}$. We can then write the state as,
\begin{eqnarray}
\label{eqn:3}
|\psi\rangle_{2,3,4}&=&\frac{1}{\sqrt{2}}[a(|\sqrt{2}\alpha,0,H\rangle_{2,3,4}+|0,\sqrt{2}\alpha,V\rangle_{2,3,4})
\nonumber \\ &&
+b(|0,-\sqrt{2}\alpha,H\rangle_{2,3,4}+|-\sqrt{2}\alpha,0,V\rangle_{2,3,4})].
\end{eqnarray}
Alice performs photon counting measurement on the output modes $3$ and $4$. The possible outcomes are $(0,0)$, $(EVEN,0)$, $(0,EVEN)$, $(ODD,0)$, $(0,ODD)$, which we enumerate by cases (i) to (v). For case (i), the state (teleported) with Bob is $|T^{0}\rangle=|H\rangle+|V\rangle$ which cannot be converted to information by any possible unitary operation, thus, leading to failure. However, its probability of occurrence, $P_{0}=\frac{x^{2}|a+b|^{2}}{1+x^{2}}$, becomes negligibly small for appreciable coherent amplitude.

For case (ii), the teleported state is $|T^{ii}\rangle=a|H\rangle+b|V\rangle$ which is exactly the information state encoded in DV basis. For cases (iii), (iv) and (v) the teleported state are, 
 $|T^{iii}\rangle=a|V\rangle+b|H\rangle$, 
 $|T^{iv}\rangle=a|H\rangle-b|V\rangle$ and 
 $|T^{v}\rangle=a|H\rangle+b|V\rangle$, respectively, requiring unitary transformations $\hat{U}^{iii}=|H\rangle\langle V|+|V\rangle\langle H|$, $\hat{U}^{iv}=|H\rangle\langle H|-|V\rangle\langle V|$ and $\hat{U}^{v}=|H\rangle\langle V|-|V\rangle\langle H|$ for respectively, so as to get the information state.
  
We, therefore, conclude that for all non-zero photon count cases perfect quantum teleportation from CV to DV can be achieved. Additionally, assuming the involved coherent amplitudes to be appreciably large, the occurrence of case (i) becomes negligibly small.
\section{Quantum teleportation from DV to CV encoding}
\label{sec:1}
We shall now discuss the quantum teleportation scheme from DV to CV encodings. Unlike teleportation from CV to DV, achieving perfect teleportation for this case is not trivial. Park et al. \cite{park2012} proposed a scheme to obtain quantum teleportation from DV to CV encoding using hybrid-entangled resource with maximum success probability of $1/2$. The reduced success is due to the fact that in their scheme the sender is required to perform Bell state measurement on DV qubits. It is well known fact that using linear optics we can only discriminate two out four Bell states \cite{lutkenhaus1999bell}. Another reason for failure is the non unitary nature of $\hat{Z}_c=|\alpha\rangle\langle-\alpha|-|-\alpha\rangle\langle\alpha|$ gate-operation for coherent state qubit. Therefore, using hybrid-entangled resource of the kind given by Eq.~\ref{eqn:2}, quantum teleportation from DV to CV system cannot be obtained with success greater than $1/2$. 
\subsection{Time evolution in a cross-Kerr interaction}
As we shall be using cross-Kerr non linear interaction we briefly discuss its dynamics at this point.
The interaction Hamiltonian for cross-Kerr non-linearity is given by, 
$\hat{H}_{\mathrm{ck}}=\hslash\chi\hat{N}_{x}\hat{N}_{y}$, where $\chi$ is proportional to strength of third-order non-linearity and $\hat{N}_{x,y}$ are the number operator for input modes $x$ and $y$. Within such a medium, the interaction between input modes is governed by unitary evolution,
\begin{equation}
\label{eqn:4}
\hat{U}_{ck}=\exp[-it\chi\hat{N}_{x}\hat{N}_{y}],
\end{equation}
where $t$ is the interaction time. Assuming one of the input mode to be a coherent state $|\alpha\rangle$ and another Fock state $|n\rangle$ then after a given time $t$ the states would evolve as,
\begin{equation}
\label{eqn:5}
\hat{U}_{ck}|\alpha\rangle_{x}|n\rangle_{y}=|\alpha e^{-in\theta}\rangle_{u}|n\rangle_{v},
\end{equation}
where $\theta=\chi t$. It is interesting to note that the Fock state remains unchanged while the coherent state picks a phase which is proportional to the number of photons in the Fock state. It is this property of cross-Kerr interaction that makes it important for many quantum information processing tasks as well as for making quantum non-demolition measurement of photon number \cite{munro2005,gerry2008}. We shall also use this property to entangle an auxiliary coherent state with the polarization qubit.
\begin{figure}
\includegraphics[width=\textwidth]{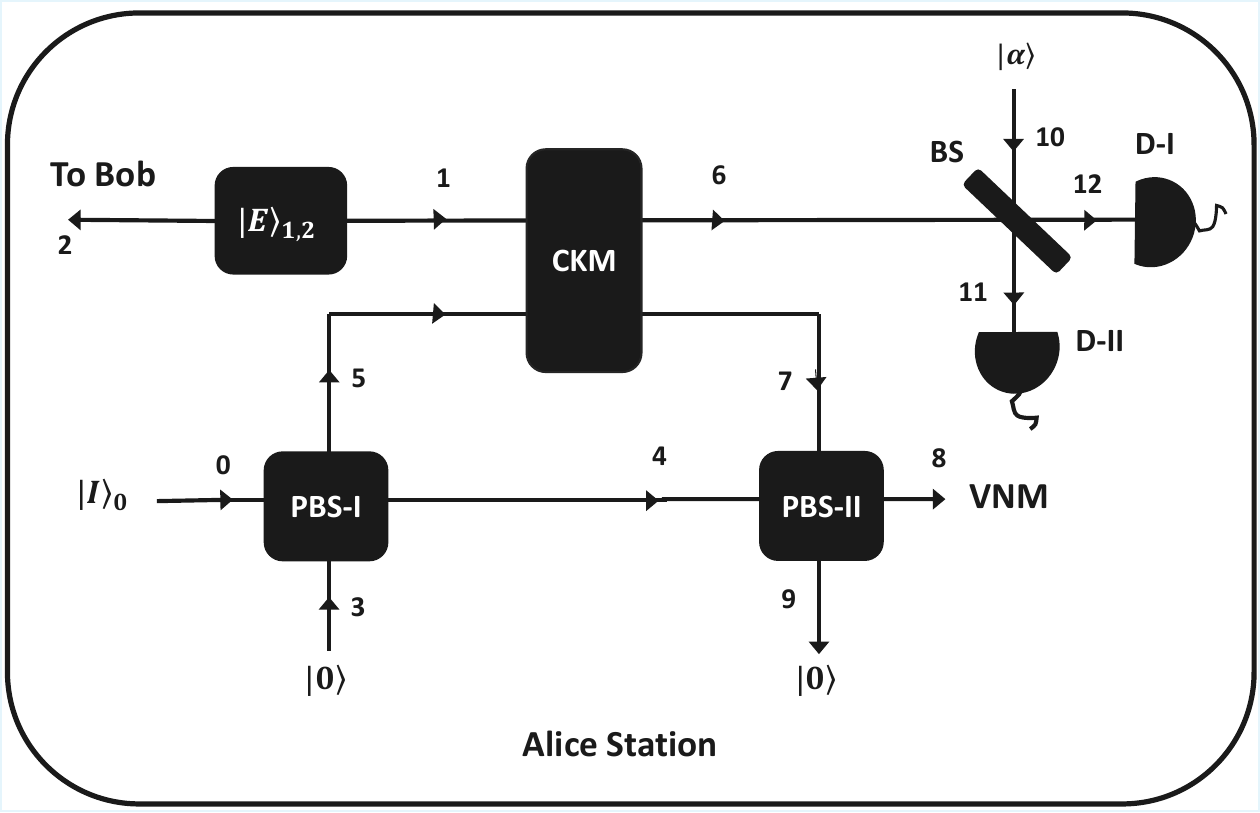}
    \vfill
    \caption{Schematic of the proposed scheme for quantum teleportation from DV to CV quantum system. $|E\rangle_{1,2}$ is the shared entangled resource where mode 1 is with Alice and mode 2 is sent to Bob. Alice initially has DV information state in mode 0 which she passes through a polarizing beam splitter (PBS-I) to become mode 3. Modes 1 and 3 is passed through a cross Kerr medium (CKM) to give output modes 6 and 5. Alice then passes modes 4 and 5 through another PBS-II that outputs modes 7 and 8. VNM is performed on mode 7 while mode 6 is either $|\alpha\rangle$ or $|-\alpha\rangle$ which is further discriminated by the combination of a symmetric beam splitter (BS) and two photon counting detectors D-I and D-II. }
    \label{fig2}
\end{figure}
\subsection{Teleportation scheme}
Let Alice has an information state encoded in DV polarization basis in mode 0
\begin{equation}
\label{eqn:6}
|I\rangle_{0}=a|H\rangle_{0}+b|V\rangle_{0},
\end{equation}
and Bob is interested to obtain the information in CV encoding. Instead of sharing the hybrid-entangled channel of the form given by Eq.~\ref{eqn:2}, Alice and Bob share modes $1$ and $2$ of maximally entangled coherent state,
\begin{equation}
\label{eqn:7}
|E\rangle_{1,2}=[2(1-x^{4})]^{-\frac{1}{2}}(|\alpha,\alpha\rangle_{1,2}-|-\alpha,-\alpha\rangle_{1,2}),
\end{equation} 
respectively. 

Alice passes mode $0$ through a polarizing beam splitter (PBS-I), that transmits horizontally polarized photon and reflects a vertically polarized photon. This will transform the information state $|I\rangle_0$ to $a|H,0\rangle_{4,5}+b|0,V\rangle_{4,5}$. She then passes modes $1$ and $5$ into a cross-Kerr medium to get output modes $6$, $7$. The interaction time is so chosen that a phase of $\pi$ is induced in the coherent state during its interaction with $|V\rangle$. Alice recombines modes $7$ and $4$ by passing it through another polarizing beam splitter (PBS-II) which outputs modes $9$ and $8$. Dropping mode $9$ which is a vacuum, the state of the system in modes $8$, $6$ and $2$ becomes,
\begin{equation}
\label{eqn:8}
|\psi\rangle_{8,6,2}=[2(1-x^{4})]^{-\frac{1}{2}}[a|H,\alpha,\alpha\rangle-a|H,-\alpha,-\alpha\rangle_{8,6,2}+b|V,-\alpha,\alpha\rangle-b|V,\alpha,-\alpha\rangle_{8,6,2}].
\end{equation} 
Alice mixes mode $6$ with an auxiliary coherent state $|\alpha\rangle$ in mode $10$ using symmetric beam splitter that outputs modes $11$ and $12$. We expand $|\pm\sqrt{2}\alpha\rangle=\sqrt{1-x^{2}}|NZ\rangle+x|0\rangle$, where $|NZ\rangle$ contains only nonzero Fock states. The state of the system consisting of modes $8, 11, 12$ and $2$ will become,
\begin{eqnarray}
\label{eqn:9}
|\psi\rangle_{8,11,12,2}
&=&
\frac{1}{2\sqrt{1+x^{2}}}
\Big\{
|+\rangle_{8}|NZ\rangle_{11}|0\rangle_{12}
\left(a|\alpha\rangle_{2}-b|-\alpha\rangle_{2}\right)
\nonumber\\
&&
+|-\rangle_{8}|NZ\rangle_{11}|0\rangle_{12}
\left(a|\alpha\rangle_{2}+b|-\alpha\rangle_{2}\right)
\nonumber\\
&&
+|+\rangle_{8}|0\rangle_{12}|NZ\rangle_{11}
\left(-a|-\alpha\rangle_{2}+b|\alpha\rangle_{2}\right)
\nonumber\\
&&
+|-\rangle_{8}|0\rangle_{12}|NZ\rangle_{11}
\left(-a|-\alpha\rangle_{2}-b|\alpha\rangle_{2}\right)
\nonumber\\
&&
+\sqrt{2}x\,|+\rangle_{8}|0\rangle_{12}|0\rangle_{11}
(a+b)|ODD;\alpha\rangle_{2}
\nonumber\\
&&
+\sqrt{2}x\,|-\rangle_{8}|0\rangle_{12}|0\rangle_{11}
(a-b)|ODD;\alpha\rangle_{2}
\Big\}.
\end{eqnarray}
where $|ODD;\alpha\rangle=\frac{|\alpha\rangle-|-\alpha\rangle}{\sqrt{2(1-x^{2})}}$ is the odd coherent state and $|\pm\rangle=[2]^{-\frac{1}{2}}[|H\rangle \pm |V\rangle]$.
Alice performs von-Neumann measurement (VNM) on mode 8 in $|\pm\rangle$ basis at the same time she also find out whether mode 6 is $|0\rangle$ or $|NZ\rangle$ using photon detectors $D-I$ and $D-II$. The measurement outcomes obtained by Alice is discussed in the next section.
\subsection{Von-Neumann measurement and photon detection outcomes}

\begin{table*}[t]
\centering
\resizebox{\textwidth}{!}{%
\begin{tabular}{c c c c c c c}
\hline
\\
Cases &
\multicolumn{3}{c}{Alice's Measurement Result} &
Bob's State (mode 2) &
Required Unitary (Bob) &
Fidelity
\\
& Mode 8 & Mode 11 & Mode 12 & & & \\
\hline
\\
(i)  & $|+\rangle$ & $|NZ\rangle$ & $|0\rangle$
& $a|\alpha\rangle-b|-\alpha\rangle$
& $\hat{D}(\delta)$
& $F$
\\[2ex]

(ii) & $|-\rangle$ & $|NZ\rangle$ & $|0\rangle$
& $a|\alpha\rangle+b|-\alpha\rangle$
& $\hat{I}$
& $1$
\\[2ex]

(iii) & $|+\rangle$ & $|0\rangle$ & $|NZ\rangle$
& $a|-\alpha\rangle-b|\alpha\rangle$
& $\hat{D}(\delta)\hat{P}(\pi)$
& $F$
\\[2ex]

(iv) & $|-\rangle$ & $|0\rangle$ & $|NZ\rangle$
& $a|-\alpha\rangle+b|\alpha\rangle$
& $\hat{P}(\pi)$
& $1$
\\[2ex]

(v) & $|+\rangle$ & $|0\rangle$ & $|0\rangle$
& $|ODD;\alpha\rangle$
& --
& $F_{0}$
\\[2ex]

(vi) & $|-\rangle$ & $|0\rangle$ & $|0\rangle$
& $|ODD;\alpha\rangle$
& --
& $F_{0}$
\\[2ex]

\hline
\end{tabular}%
}
\caption{Results of the von-Neumann measurement performed by Alice on mode 8 together with photon
counting on modes 11 and 12. Cases (ii) and (iv) yield perfect teleportation with unit fidelity,
cases (i) and (iii) lead to approximate teleportation with fidelity $F$ given by
Eq.~(\ref{eqn:14}), while cases (v) and (vi) correspond to failure events where Bob receives an odd
coherent state carrying no information about the input qubit. $\hat{D}(\delta)$ and $\hat{P}(\pi)$ denote the displacement operator and the $\pi$ phase-shift operation, respectively.}
\label{t1}
\end{table*}
The von-Neumann measurement (VNM) performed by Alice on mode 8, together with photon detection on
modes 11 and 12, leads to six mutually exclusive measurement outcomes, as summarized in
Table~\ref{t1}. Among these, cases (ii) and (iv) correspond to successful teleportation with unit
fidelity. In particular, for case (ii) the conditional state received by Bob in mode 2 is exactly the
CV information state and no unitary transformation is required. For case (iv), Bob needs to apply a
$\pi$ phase shift operation $\hat{P}(\pi)$ on mode 2 in order to recover the information state.

The problem with unitary correction mainly arises for cases (i) and (iii). As discussed earlier, the operator $\hat{Z}_c$ is non-unitary, and therefore its application
results in irreversible loss of quantum information. This difficulty can be circumvented by adopting
the scheme proposed by Pandey \textit{et al.}~\cite{pandey2021}, in which the receiver is required to
apply only a $\pi$ phase shift to obtain CV teleportation with near-unit success probability. The same
approach can be employed here as well.

An alternative solution can be obtained by following the method proposed by Jeong
\textit{et al.}~\cite{jeong2001quantum}, where the required $\hat{Z}_c$ operation is approximated by
displacing the coherent-state qubit. A coherent state can be expressed as a displaced vacuum state,
\begin{equation}
|\alpha\rangle \equiv \hat{D}(\alpha)|0\rangle ,
\label{eqn:10}
\end{equation}
where the displacement operator is defined as
$\hat{D}(\alpha)=\exp(\alpha a^{\dagger}-\alpha^{*}a)$.
In general, two displacement operators do not commute, and two successive displacements by
$\alpha$ and $\beta$ differ from a single displacement $(\alpha+\beta)$ by a phase factor. More
precisely,
\begin{equation}
\hat{D}(\beta)\hat{D}(\alpha)
=
\exp\!\left[\tfrac{1}{2}(\alpha\beta^{*}-\alpha^{*}\beta)\right]
\hat{D}(\alpha+\beta).
\label{eqn:11}
\end{equation}

Let us consider case (i), where the conditional state received by Bob is
\(
|I'\rangle_2=a|\alpha\rangle_2-b|-\alpha\rangle_2
\).
By applying a displacement operation $\hat{D}(\delta)$, Bob obtains
\begin{equation}
\hat{D}(\delta)|I'\rangle_2
\sim
\left[
a\hat{D}(\alpha+\delta)
-
b\,e^{(\alpha\delta^{*}-\alpha^{*}\delta)}\hat{D}(-\alpha+\delta)
\right]|0\rangle_2 .
\label{eqn:12}
\end{equation}
For real $\alpha$ and $\delta=i\gamma$, this simplifies to
\begin{equation}
\hat{D}(\delta)|I'\rangle_2
\sim
\left[
a\hat{D}(\alpha+\delta)
-
b\,e^{-2i\gamma\alpha}\hat{D}(-\alpha+\delta)
\right]|0\rangle_2 .
\label{eqn:13}
\end{equation}
Choosing $\gamma\alpha=\pi/4$ the teleported
state becomes
\(
|T\rangle_2\sim a|\alpha+\delta\rangle_2+b|-\alpha+\delta\rangle_2
\).
If we choose the appropriate reference state to be
\(
|I_\delta\rangle=a|\alpha+\delta\rangle+b|-\alpha+\delta\rangle
\),
then teleportation fidelity for this case becomes
\begin{equation}
F
=
|\langle I_\delta | T \rangle|^2
=
e^{-|\delta|^2}
\left[
1+2x^2\mathrm{Re}(a^*b)
\right]
\label{eqn:14}
\end{equation}
which approaches unity for appreciable coherent amplitude
$|\alpha|^2$.
We note that for case (iii) Bob has to perform a $\pi$ phase shift and then $\hat{D}(\delta)$ to obtain the teleported state $|I_\delta\rangle$, resulting in similar value of fidelity as for case (i).

Cases (v) and (vi) correspond to simultaneous vacuum detection in both modes~11 and~12. In these
events, the state received by Bob collapses to the odd coherent state
$|ODD;\alpha\rangle$
which is independent of the information parameters $a$ and $b$. These outcomes therefore
represent failure events. For completeness, the associated fidelity is given by
$
F_{0}
=
(1-x^2)|a-b|^2/2$,
which has a maximum value of $1/2$.
\subsection{Average fidelity of teleportation}
\label{sec:avg_fidelity}

\begin{figure}
\begin{center}
\includegraphics[width=0.8\textwidth]{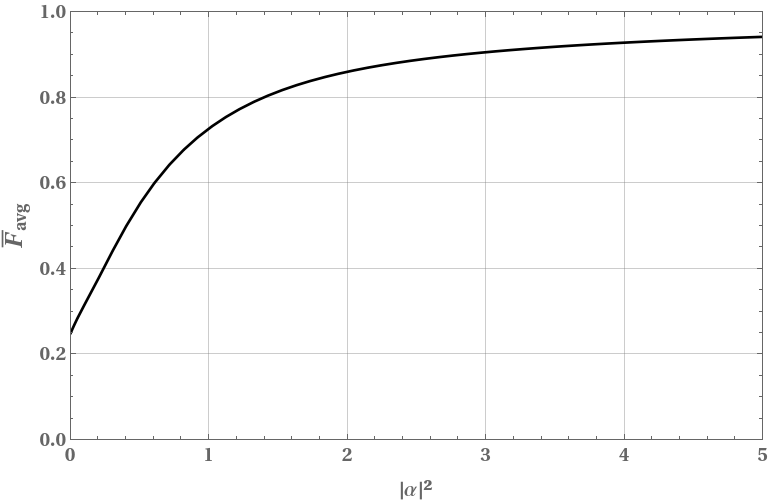}
    \vfill
    \caption{Variation of average fidelity over information parameters $a$ and $b$ of teleportation for DV to CV encodings. The fidelity increase monotonically and sharply to approach unity asymptotically for appreciable large coherent amplitude.}
    \label{fig3}
    \end{center}
\end{figure}

In order to quantify overall quality of teleportation protocol, we shall find average fidelity of the DV-CV teleportation protocol. The average fidelity is defined as a probability-weighted sum of
the fidelities corresponding to individual measurement outcomes. The probability of occurrence of various cases can be directly obtained using Eq.~\ref{eqn:9}. Explicitly, we have,
\begin{equation}
\label{eqn:16}
P_i=P_{iii} = \frac{1-2x^2\mathrm{Re}(a^*b)}{4(1+x^{2})}
\end{equation}
\begin{equation}
\label{eqn:17}
    P_{ii}=P_{iv}=\frac{1+2x^2\mathrm{Re}(a^*b)}{4(1+x^{2})},
\end{equation}
\begin{equation}
\label{eqn:18}
P_{v}=\frac{x^2|a+b|^2}{4(1+x^2)}, P_{vi}=\frac{x^2|a-b|^2}{4(1+x^2)}.
\end{equation}
It is straightforward to verify that these probabilities satisfy 
$\sum_{k=i}^{vi}P_{k}=1$.
The average teleportation fidelity is now obtained as the probability-weighted sum of the individual
fidelities,
\begin{equation}
\label{eqn:19}
F_{avg}
=
2P_{ii}
+ 2P_{i}F
+ (P_{v} + P_{vi})F_{0}.
\end{equation}
Substituting the values of probabilities and fidelities using Eqs.~\ref{eqn:14}-\ref{eqn:18}, we obtain
\begin{equation}
\label{eqn:20}
F_{avg}
=
\frac{1}{2(1+x^2)}
\left\{
1
+ e^{-|\delta|^{2}}
[
1 - \big(2x^2\mathrm{Re}(a^*b)\big)^2
]+ 2x^2\mathrm{Re}(a^*b)
+ x^2(1-x^2)|a-b|^2
\right\}.
\end{equation}
We note that the average fidelity is a function of the information parameters $a$ and $b$ as well as the mean coherent amplitude $|\alpha|^2$. To assess the overall performance of the protocol that is independent of a specific input state, we now
average Eq.(\ref{eqn:20}) uniformly over the Bloch sphere. A general input qubit is
parameterized as
$a=\cos\frac{\theta}{2}$,
$b=e^{i\phi}\sin\frac{\theta}{2}$ 
with $\theta\in[0,\pi]$ and $\phi\in[0,2\pi]$. Performing the integral over the Bloch-sphere, the average state fidelity over the information parameters ($\overline{F}_{avg}$) is given by,
\begin{equation}
\label{eqn:21}
\overline{F}_{avg}
=
\frac{1}{2(1+x^2)}
\left\{1
+ e^{-|\delta|^{2}}\left(1-\frac{x^4}{3}\right)
+ x^2(1-x^2)\right\}.
\end{equation}
Fig.~\ref{fig3} shows the monotonic increase of the average state fidelity with respect to $|\alpha|^2$. Specifically, for $|\alpha|^2=5$, the fidelity reaches 0.932, and it approaches unity asymptotically as the coherent amplitude becomes large ($|\alpha|^2\gg0$). In this macroscopic limit, where $x\rightarrow0$ and $|\delta|^2\rightarrow1$, the protocol becomes deterministic. These results demonstrate that the proposed scheme provides an efficient and near-perfect mechanism for quantum teleportation between DV and CV systems.
\section{Conclusion}
We conclude that our present scheme is capable of obtaining quantum teleportation between DV and CV encodings of quantum information with success probability almost unity. In principle, teleportation from CV to DV quantum system is straight- forward and can be achieved using linear optical elements such as beam splitters, phase shifters, and photon counters. On the contrary, teleportation from  DV to CV quantum systems poses two major limitation: firstly, using linear optical elements the Bell state measurement on DV polarization qubits can discriminate only two out of four Bell states, and secondly, for certain Bell state outcomes, the receiver needs to apply $\hat{Z}_c$ operator which is not unitary. In order to circumvent these issues, we use cross-Kerr non-linearity to entangle the polarization qubit with the one of the modes of an entangled coherent state. The Bell state measurement is then performed on CV coherent modes which can be achieved with unit success for appreciable coherent amplitude. The issue of performing non-unitary $\hat{Z}_c$ can be overcome by displacing the CV qubit by a value $\delta$ which depends inversely on the coherent amplitude $\alpha$. Therefore, for appreciably large $\alpha$, the displacement $\delta\rightarrow 0$ and the fidelity of teleported state becomes almost unity. 

In a realistic scenario, the optical modes traveling through the quantum channel tend to attenuate, amounting to a loss in entanglement \cite{zhang2021}. Therefore, it is important to study the effect of such losses in the present scheme also, and this shall be a part of our future investigations. Furthermore, our scheme can be used to obtain almost perfect hybrid quantum information processing tasks such as entanglement diversion, remote state preparation, superdense coding etc., and for other variants of hybrid quantum teleportation protocols such as controlled quantum teleportation, bidirectional teleportation and cyclic quantum teleportation. 

\medskip
\noindent
\textbf{Author contribution} The authors completed the research and writing of this paper together. All authors
reviewed the manuscript.

\medskip
\noindent
\textbf{Funding} RKP would like to acknowledge the Council of Scientific $\&$ Industrial Research (CSIR),
New Delhi, India, for Research Associate Fellowship. DKM acknowledges the Institution of Eminence (IoE), Banaras Hindu University, Varanasi, India for the Bridge Grant.

 \bibliographystyle{spphys}
 \bibliography{QTrevised}  

\end{document}